\documentclass[aps,prb,reprint,floatfix]{revtex4-1}
\usepackage{epsfig}
\usepackage{amsmath}
\usepackage{hyphenat}
\usepackage{hyperref}
\usepackage{placeins}
\usepackage[justification=raggedright]{caption}	
\usepackage[colorinlistoftodos]{todonotes}

\begin{document}

\title{The skyrmion lattice phase in three dimensional chiral magnets from Monte Carlo simulations}
\author{Stefan Buhrandt and Lars Fritz}
\affiliation{Institute for Theoretical Physics, University of Cologne, Cologne, Germany}

\begin{abstract}
Chiral magnets, such as MnSi, display a rich finite temperature phase diagram in an applied magnetic field. The most unusual of the phases encountered is the so called A-phase characterized by a triangular lattice of skyrmion tubes. Its existence cannot be captured within a mean-field treatment of a Landau-Ginzburg functional but thermal fluctuations to Gaussian order are required to stabilize it~\cite{Muehlbauer}. In this note we go beyond Gaussian order in a fully non-perturbative study of a three dimensional lattice spin model using classical Monte Carlo simulations. We demonstrate that the A-phase is indeed stabilized by thermal fluctuations and furthermore we reproduce the full phase diagram found in experiment. The thermodynamic signatures of the helimagnetic transition upon cooling from the paramagnet are qualitatively consistent with experimental findings and lend further support to the Brazovskii scenario~\cite{Brazovskii1975} which describes a fluctuation driven first order transition due to the abundance of soft modes~\cite{papermarkus}. 
\end{abstract}
\maketitle

\section{Introduction}

Chiral magnets like MnSi or Fe$_{1-x}$Co$_x$Si have received a lot of interest recently~\cite{Muehlbauer,Muenzer,Nature2DSkyrmions,NatureMaterials}, mainly by virtue of them showing a thermodynamic phase which is characterized by a lattice consisting of tubes of magnetic skyrmions. Besides the very existence of this phase there seems to be huge potential to use these materials for spintronics applications~\cite{Schulz12}. 

The lack of inversion symmetry in the crystalline structure of these magnets gives rise to weak Dzyaloshinskii-Moriya (DM) coupling. The competition of this interaction with the much stronger ferromagnetic exchange (FM) results in a twist in the magnetic order, leading to helical order. Since the DM coupling is weak compared to the FM exchange coupling there are long modulation periods of many lattice constants, e.\,g. in the chiral magnet prototype MnSi the modulation period is about $190\,$\AA\, while the lattice constant is only $4.6\,$\AA \cite{Muehlbauer}. The competition between these two types of interactions determines the length of the magnetic spirals but not their direction. Consequently, one expects a large ground state degeneracy at zero magnetic field. This degeneracy is, however, lifted by weak crystal anisotropies which provide an easy axis for the ordering wave vector (e.\,g. [111] in MnSi). As a direct consequence the phase with helical order has a single ordering wave vector. If additionally a finite magnetic field is applied it becomes energetically favorable to have the ordering wave vector point in direction of the magnetic field. All spins then point in a plane perpendicular to the field and the system can gain energy by simply tilting all spins continuously out of that plane in direction of the field, leading to a spiralling umbrella structure. This state is referred to as conical phase. Depending on the direction of the field the phase transition between these two phases is either first order or a crossover and occurs at some field value $B_{c1}$ where the energy gain from tilting all spins towards the field becomes larger than the crystal anisotropy energy difference between the two directions of the ordering wave vector. Figs. \ref{fig:sf} a.) and b.) schematically show the magnetization structure in these two phases.

In 2009 neutron refraction experiments on MnSi\cite{Muehlbauer} discovered a new thermodynamic phase at intermediate fields and temperatures just below $T_c \approx 30$\,K. This phase is characterized by a periodic arrangement of tubes of skyrmions which arrange on a triangular lattice (see Fig. \ref{fig:sf} c.)). Consequently, this phase is referred to as skyrmion lattice phase or A-phase. The skyrmion lattice phase can be pictured as a superposition of three helices with equal pitch length and relative angle of 120 degrees in the plane perpendicular to the magnetic field. While in mean-field theories based on a minimal Landau-Ginzburg theory for anisotropic non-centrosymmetric magnets the skyrmion lattice was argued to be a stable solution, this phase does not appear as a stable phase and always is slightly higher in its free energy than the conical phase for cubic systems~\cite{Bogdanov1,Bogdanov2}, such as MnSi. While it was argued that this phase could still be stabilized by long-ranged interactions~\cite{Tewari,Fischer} or extra phenomenological parameters~\cite{Roessler} in the free energy functional, M\"uhlbauer \cite{Muehlbauer} et al. found that a very natural alternative mechanism to stabilize the skyrmion phase is given by thermal fluctuations to a Gaussian level on top of the mean-field theory. 
 
In order to make this argument stronger it is desirable to use an approach which is not based on Gaussian fluctuations, but instead incorporates the thermal fluctuations in a non-perturbative manner, namely classical Monte Carlo (MC) simulations. Simulations for two dimensional systems \cite{PhysRevB.80.054416,Nature2DSkyrmions} have been performed before and indeed found a stable skyrmion lattice phase. The phase diagrams obtained are in excellent agreement with recent experiments \cite{Nature2DSkyrmions} on thin films of Fe$_{0.5}$Co$_{0.5}$Si even though the itinerant character of the underlying electronic system is not taken into account in these studies. A major reason why these studies have not been extended to three dimensional systems yet is the high computational effort: large system sizes are required to account for the long spatial modulations. A further complication stems from the fact that one has to be very careful in choosing effective parameters of the underlying lattice-spin model.

In this paper we fill this void and perform a MC study for three dimensional chiral magnets. We confirm that the effect of thermal fluctuations indeed is what stabilizes the skyrmion phase~\cite{Muehlbauer}. Overall, we find excellent agreement with the experimentally observed phase diagram as well as with non-trivial thermodynamic signatures across the phase boundary from the paramagnet into one of the respective ordered phases. For zero magnetic field the transition from the paramagnet into the helimagnet is a fluctuation-driven first order transition and can be described in terms of the Brazovskii scenario~\cite{Brazovskii1975}. Some of the experimental features of this transition have proven hard to capture in purely analytical approaches~\cite{papermarkus}, however, the Monte Carlo approach captures all the qualitative features.

The organization of the paper is as follows: We start with a discussion of the model and the method. Most importantly, we introduce a minimal lattice model which is consistent with the system symmetries and consequently the Ginzburg-Landau functional. We furthermore introduce the Metropolis algorithm together with the algorithm which is required to overcome the large hysteresis in the underlying system. From there we move to the global phase diagram of a chiral magnet in an applied magnetic field and compare it to experimental findings. We close with a comparison of some thermodynamic quantities to the experimental findings in zero and non-zero field as we go across the thermal transition from the high temperature paramagnet towards one of the ordered phases. We find excellent agreement with experiment which lends further support to the relevance of our approach to this problem, despite the rather small lattice sizes.

\section{Model \& methods} 
\label{sec:model}

\subsection{The lattice Hamiltonian}

Assuming a slow variation of the spin textures one can resort to a coarse grained continuum model for the description of the magnetic properties of chiral magnets. The commonly used one assumes the form (cf.\cite{0022-3719-13-31-002,Nature2DSkyrmions})
{\allowdisplaybreaks
\begin{multline}
		H = \int \text{d}^3\mathbf{r}  \Bigg[  \frac{J}{2a} \left( \nabla \mathbf{M(r)} \right)^2 - \frac{\mathbf{B}\cdot\mathbf{M(r)}}{a^3} \\
			+\frac{K}{a^2} \mathbf{M(r)} \cdot \left( \nabla \times \mathbf{M(r)} \right) \Bigg],
\label{eq:ContinuumHamiltonian}
\end{multline}}
consisting of ferromagnetic exchange $J$, magnetic field $\mathbf{B}$, and a DM interaction $K$. Above, $a$ is the typical distance over which the spin structure can be treated as uniformly ordered allowing for the coarse graining procedure. This effective model has to be understood in connection with the renormalization group meaning that terms accounting for the actual microscopic lattice structure can be dropped by virtue of them being irrelevant at the critical point. They can, however, be important for a faithful description deep inside the ordered phase. Instead of the full B20 structure of MnSi one compactifies the above continuum theory onto a simple cubic lattice (which in principle has inversion symmetry unless explicitly broken as we do below). The construction principle is that the effective low-energy theory which can be derived from the lattice Hamiltonian agrees with the above model up to terms which are irrelevant in the renormalization group sense.  

As a microscopic model we adopt the lattice Hamiltonian presented in \cite{PhysRevB.80.054416},

\begin{eqnarray}
H &=& -J \sum_{\mathbf{r}} \mathbf{S}_{\mathbf{r}} \cdot \left( \mathbf{S}_{\mathbf{r}+{\mathbf{\hat{x}}}} + \mathbf{S}_{\mathbf{r}+{\mathbf{\hat{y}}}} + \mathbf{S}_{\mathbf{r}+{\mathbf{\hat{z }}}} \right) -  \mathbf{B}\cdot \sum_{\mathbf{r}} \mathbf{S}_{\mathbf{r}}\nonumber \\ 
			&&-  K \sum_{\mathbf{r}} \left( \mathbf{S}_{\mathbf{r}} \times \mathbf{S}_{\mathbf{r}+{\mathbf{\hat{x}}}} \cdot \mathbf{\hat{x}} \right. \nonumber \\  && \left.+  \mathbf{S}_{\mathbf{r}} \times \mathbf{S}_{\mathbf{r}+{\mathbf{\hat{y}}}} \cdot \mathbf{\hat{y}}   +    \mathbf{S}_{\mathbf{r}} \times \mathbf{S}_{\mathbf{r}+{\mathbf{\hat{z}}}} \cdot \mathbf{\hat{z}}      \right)\;.
\label{eq:LatticeHamiltonian}
\end{eqnarray}

In the following we discuss how to extend this model in order to get rid of discretization errors which turn out to be large and decisive in the cases considered below.

\subsection{Finite size effects and anisotropies}
\label{sec:anisotropy}

The pitch length of the helices is determined by the ratio $K/J$. We choose $K/J~=~\tan(2\pi/10)\approx 0.727$ to obtain a pitch length of 10 lattice sites for a helix 
propagating in [100] direction at zero field \cite{PhysRevB.80.054416}. We have found that the maximal lattice size tractable in reasonable CPU time is given by $N=30^3$ spins, which already hosts up to nine skyrmion tubes in total (However, for isolated cases we have checked our results against simulations on lattices of size $N=40^3$ with agreeing results). We use periodic boundary conditions since open boundary conditions lead to polarized spins on the boundaries due to missing next neighbor FM and DM interaction which makes them profit maximally from the Zeeman energy. Since it is impossible to choose parameters such that helices e.\,g. in $[111]$ and $[100]$ direction fit perfectly on the lattice at the same time one would expect strong finite size effects. However, we found that this turned out not to be a major complication in our simulations. 

The discretization of the continuum model, on the other hand, creates anisotropies which have to be taken seriously. This can be seen as follows: On the lattice, the FM Heisenberg term in Eq.~\eqref{eq:LatticeHamiltonian} after Fourier transform reads
\begin{eqnarray}\label{eq:HFM}
	H_{\text{FM}} = J \sum_{\mathbf{k}} \alpha_{\mathbf{k}} \mathbf{S}(\mathbf{k}) \cdot \mathbf{S}(\mathbf{-k})\;,
\end{eqnarray}
where 
\begin{eqnarray}
	\alpha_{\mathbf{k}} &=&  -\left(\cos(k_x a) + \cos(k_y a) + \cos(k_z a) \right) \nonumber \\ &=& -  3 + \frac{a^2}{2} (k_x^2  + k_y^2 + k_z^2) \nonumber\\  & & - \frac{a^4}{24} (k_x^4 + k_y^4 + k_z^4) + \mathcal{O}(k^6) \;. 
\end{eqnarray}
which implies that all kinds of high orders in momentum terms are generated (the constant term only shifts the energies). If we contrast this from the Fourier transform of the first term in the comtinuum model, Eq.~\eqref{eq:ContinuumHamiltonian}, we see that there only the quadratic terms are present.
One would not be worried about the higher order terms in the series if the ordered state was described by a uniform spin texture: For instance, the critical properties of the purely ferromagnetic Heisenberg lattice model and the Landau-Ginzburg functionals are in perfect agreement with each other and the anisotropies do not play a role at all. This in general is not true, especially if the critical modes do not become soft at zero momentum, as is the case for the simple Heisenberg ferromagnet, but at a finite ordering wave vector, henceforth called ${\bf{Q}}$.
Since we use relatively small lattice sizes, $|{\bf{Q}}|a$ in general is on the order $\lesssim 1$ which is not a small number. Consequently, the contribution of the higher order terms is not negligible and spoils our analysis. In order to compensate for these induced anisotropies we add next-nearest neighbor interactions $H'$ to our Hamiltonian. These terms are chosen such that they do not break symmetries of the underlying system and give a better approximation to the continuum field theory in the sense of rendering corrections from higher orders of the expansion small. They assume the form
\begin{eqnarray}
H' &=& J' \sum_{\mathbf{r}} \mathbf{S}_{\mathbf{r}} \cdot \left( \mathbf{S}_{\mathbf{r}+{\mathbf{2\hat{x}}}} + \mathbf{S}_{\mathbf{r}+{\mathbf{2\hat{y}}}} + \mathbf{S}_{\mathbf{r}+{\mathbf{2\hat{z}}}} \right)+ \nonumber \\
		 & & K' \sum_{\mathbf{r}} \left( \mathbf{S}_{\mathbf{r}} \times \mathbf{S}_{\mathbf{r}+{\mathbf{2\hat{x}}}} \cdot \mathbf{\hat{x}} +  \mathbf{S}_{\mathbf{r}} \times \mathbf{S}_{\mathbf{r}+{\mathbf{2\hat{y}}}} \cdot \mathbf{\hat{y}}   + \right. \nonumber \\ 
		 & & \left.   \mathbf{S}_{\mathbf{r}} \times \mathbf{S}_{\mathbf{r}+{\mathbf{2\hat{z}}}} \cdot \mathbf{\hat{z}}      \right) \;.
\end{eqnarray}
The full $\alpha_{\mathbf{k}}$ of the Heisenberg term, see Eq.~\eqref{eq:HFM}, is now given by
\begin{eqnarray}
	\alpha_{\mathbf{k}} &=& - 3(J-J') + \frac{a^2}{2}\left(J-4J'\right) (k_x^2   + k_y^2 + k_z^2) \nonumber\\  & & - \frac{a^4}{24}\left(J-16J'\right)(k_x^4 + k_y^4 + k_z^4) + \mathcal{O}(k^6)\;.
\end{eqnarray}
This immediately shows that we can compensate the anisotropies to leading order by choosing $J'=J/16$. Repeating the same procedure for the DM term leads to $K'=K/8$.

Another way to think about this compensation is that the approximation of the gradient terms in the continuum model, Eq.~\eqref{eq:ContinuumHamiltonian}, solely by next neighbor interactions as in Eq.~\eqref{eq:LatticeHamiltonian} is not accurate if the spin configuration varies significantly from site to site. If we could simulate larger lattices we could use a smaller value of $K$ which in turn increases the modulation period of the helices. The spin configuration would then vary more smoothly and consequently the approximation in Eq.~\eqref{eq:LatticeHamiltonian} becomes better. To summarize, the purpose of the next-nearest neighbor interaction terms $J'$ and $K'$ is to improve the approximation of the gradient terms in Eq.~\eqref{eq:ContinuumHamiltonian} by compensating the relatively short pitches in our simulation.

\subsection{Determination of thermodynamic phases and MC algorithm}

\begin{figure*}[t]
\centering
	\includegraphics[width=0.8\textwidth]{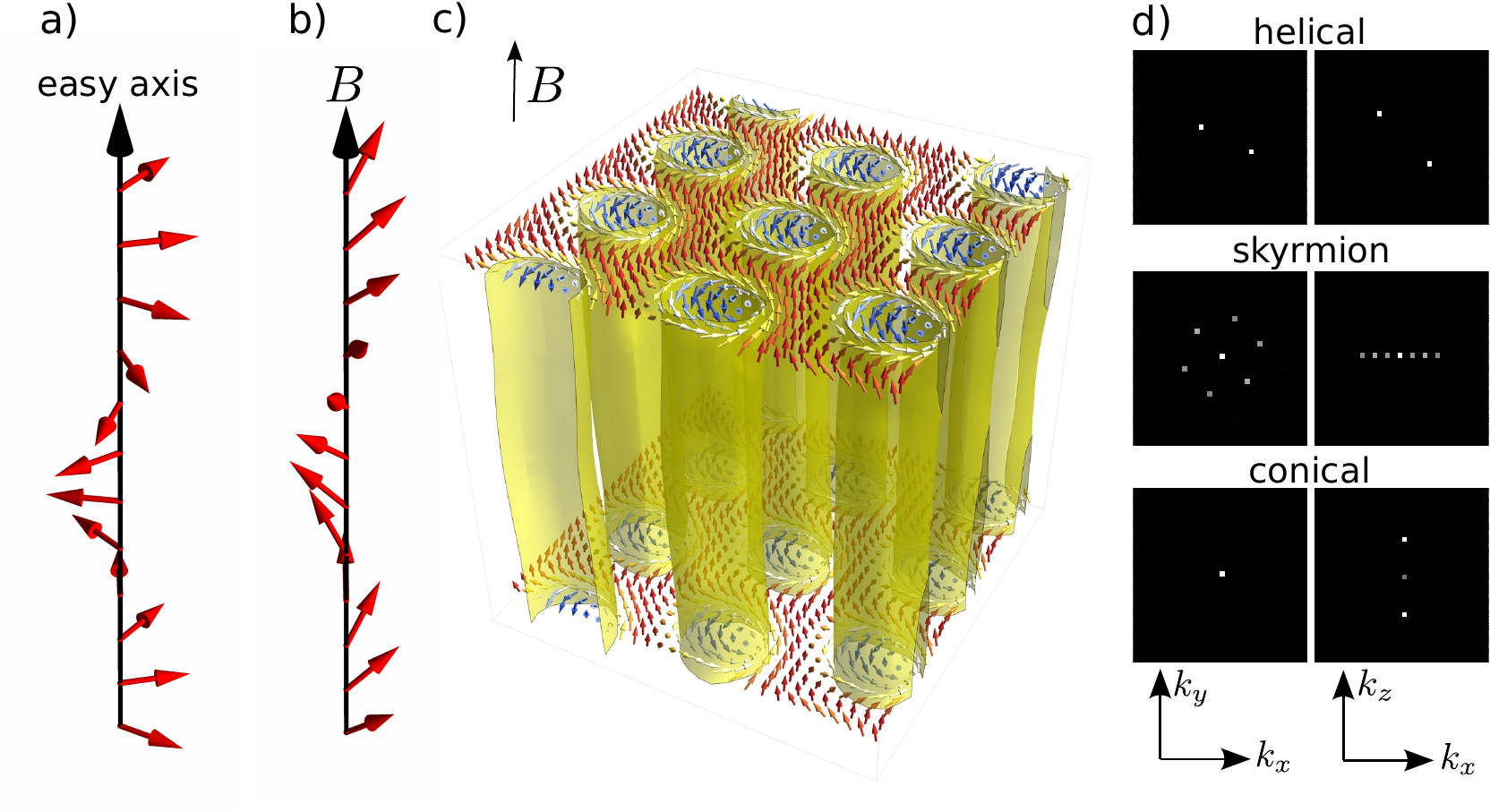}
	\caption{\textbf{a)} Schematic sketch of the magnetization in the helical phase. \textbf{b)} Schematic sketch of the magnetization in the conical phase. \textbf{c)} Averaged magnetization in the skyrmion phase in two different crystal planes for $(J,K,B,T) = (1,\tan(2\pi/10),0.18,0.82)$. The magnetization in direction of the external field vanishes along the yellow tubes. \textbf{d)} Bragg intensity patterns projected into the (001) plane (which is $\perp \mathbf{B}$) (left) and (010) plane (right). Parameters are $(J,K,B,T) = (1,\tan(2\pi/10) ,0,0)$ (helical phase), $(1,\tan(2\pi/10) ,0.18,0.82)$ (skyrmion phase) and $(1,\tan(2\pi/10) ,0.18,0)$ (conical phase).  }
	\label{fig:sf} 
\end{figure*}

The different phases in our problem can be distinguished either from the real space spin textures or, more easily, from the spin structure factor in reciprocal space. We calculate the average spin configuration $\langle~\mathbf{S}_{\mathbf{r}}~\rangle$ from usually 2000 spin configurations separated by 30 lattice sweeps and then Fourier transform the average configuration into momentum space,
\begin{equation}
	\langle~\mathbf{S}_{\mathbf{k}}~\rangle~=~\frac{1}{N}~\sum_{\mathbf{r}}~\langle~\mathbf{S}_{\mathbf{r}}~\rangle\exp(-i \,\mathbf{k\cdot r})\;.
\end{equation}
Afterwards, we analyze the Bragg intensity profile $I(\bf{k})~\propto~\|\langle~\bf{S}_{\bf{k}}~\rangle\|^2$, which corresponds to what is measured in neutron scattering experiments. A single helix with wave vector $\mathbf{Q}$ is characterized by two Bragg spots sitting at $\mathbf{Q}$ and $-\mathbf{Q}$ (as required by the real order parameter). The helical and conical phase can thus easily be distinguished by the direction of $\mathbf{Q}$ (while $\mathbf{Q}$ is parallel to the magnetic field in the conical phase it is along [111] in the helical). The skyrmion lattice phase has a richer structure and is easily identified by its six Bragg spots which are arranged on a regular hexagon in the plane perpendicular to $\mathbf{B}$. The real space spin structures for the skyrmion lattice phase as well as the Bragg intensity patterns for all three phases are shown in Fig.~\ref{fig:sf}.

We determine the spin configurations using a single-site Metropolis algorithm. The model at hand turns out to be very hysteretic, in fact we checked that even parallel tempering MC (PTMC) is not able to describe the phase transition from the skyrmion to the conical phase. In order to overcome this and ensure consistency we use three different schemes in parallel: (i) Simulated annealing, meaning cooling at constant field. (ii) Simulated annealing to the target temperature at zero field followed by slowly increasing the field. (iii) Simulated annealing to a target temperature at high field (such that we always remain in the spin polarized phase) followed by decreasing the field.
 
If no unique state is reached this means the single-site Metropolis algorithm is trapped in a metastable state. We then use a global update which allows the system to fluctuate between these states and allows to determine the true thermodynamic state: In practice, we simulate two systems in the respective different states in parallel and offer each system $2\cdot 10^5$ times to take the state of the other one after performing 20 lattice sweeps using single-site updates in its own state. The transition probabilities
\begin{equation}
	p(i\rightarrow j) = \text{min}\left(1,\exp((E_j-E_i)/T)\right)
\end{equation}
for these hypothetical steps are recorded and averaged. The detailed balance condition reads 
\begin{eqnarray}
\frac{p(i)}{p(j)}=\frac{p(j\to i)}{p(i \to j)}
\end{eqnarray}
where $p(i)$ and  $p(j)$ are the probabilities to find the system in the respective states. According to Ref.~\cite{JCC:JCC21450}, this can be used to calculate the free energy difference between the two states according to
\begin{eqnarray}
\Delta F_{ij}=-T \ln \frac{p(i)}{p(j)}\;,
\end{eqnarray}
which allows to determine the thermodynamic state by selecting the one with the lower free energy.

\section{Global phase diagram}
\label{sec:phasediagram}

\begin{figure}[h]
\centering
	\includegraphics[angle=270,width=\columnwidth]{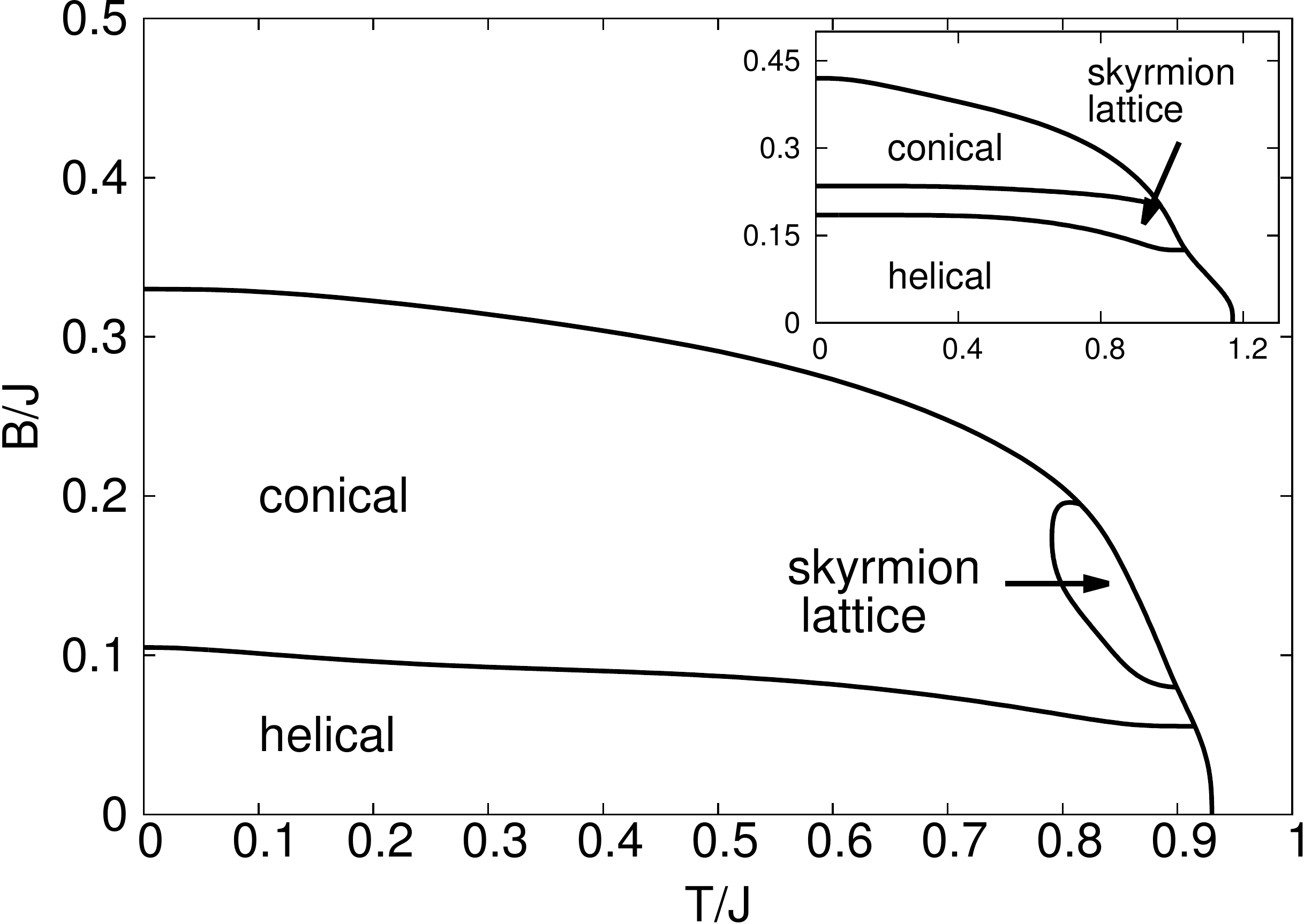}
	\caption{Phase diagram with anisotropy compensation for $(J,K)=(1,\tan(2\pi/10))$. Next-nearest neighbor interactions are chosen according to Sec.~\ref{sec:anisotropy} as $J'=J/16$ and $K'=K/8$. The inset shows the phase diagram without anisotropy compensation.}
	\label{fig:phasediagram}
\end{figure}

Our main result is the non-perturbative determination of the phase diagram associated with the free energy functional introduced in Eq.~\eqref{eq:ContinuumHamiltonian}, see Fig.~\ref{fig:phasediagram}.

As mentioned in Sec.~\ref{sec:anisotropy}, we have to account for the presence of discretization errors by subtracting the quartic terms of the nearest neighbor interaction. To show that our simulations are severely hampered by these effects, we have determined the $(B,T)$-phase diagram with and without anisotropy compensation.

In both cases our simulation finds all three ordered phases found in the experiment on MnSi (Fig.~\ref{fig:sf} for their real and reciprocal space signatures). In particular we find a helical phase for small magnetic fields even though our Hamiltonian does not provide any explicit anisotropies that favor a certain crystal direction. These anisotropies are, however, automatically generated in the lattice model due to discretization errors and seem to favor propagation in [111] direction in our case, as explicit in Fig.~\ref{fig:sf}.

Most importantly, our simulation shows a stable skyrmion phase at intermediate fields. Fig.~\ref{fig:phasediagram} shows the phase diagram obtained with anisotropy compensation. The skyrmion phase is stable only in a small pocket close to $T_c$ (c.\,f. Fig.~\ref{fig:phasediagram}) as it is also observed experimentally. Without the anisotropy compensation on the other hand, the skyrmion phase remains stable even for $T \to 0$ (c.\,f. inset in Fig.~\ref{fig:phasediagram}), which is in disagreement with experiment.  We conclude that the discretization anisotropies indeed spoil our analysis and the true phase diagram is only obtained after compensation of these effects to leading order. The real-space spin configuration of the skyrmion phase obtained from our MC simulations is shown in Fig.~\ref{fig:sf}. 

This behavior has to be contrasted from previous analysis in the case of two dimensional thin film systems. In numerical simulations for two dimensional systems or thin films with the field perpendicular to the plane one did not encounter the need for anisotropy compensation. This is related to the fact that there the conical phase ceases to exist (since in the conical phase the spin texture likes to propagate along the magnetic field) and no competition between the conical and skyrmion phase takes place. Consequently, the skyrmion phase remains stable for $T\rightarrow 0$~\cite{Nature2DSkyrmions}. 

To summarize, we reproduce the full phase diagram of three dimensional helical magnets in a non-perturbative manner which holds beyond mean-field plus low order fluctuation analysis. Our analysis conclusively shows that the original claim that thermal fluctuations lower the free energy of the skyrmion phase as compared to the conical phase within a finite pocket which was based on the lowest non-trivial order in an expansion around mean-field~\cite{Muehlbauer} is correct and higher order corrections do not spoil the picture.  
\begin{figure}
	\includegraphics[width=\columnwidth]{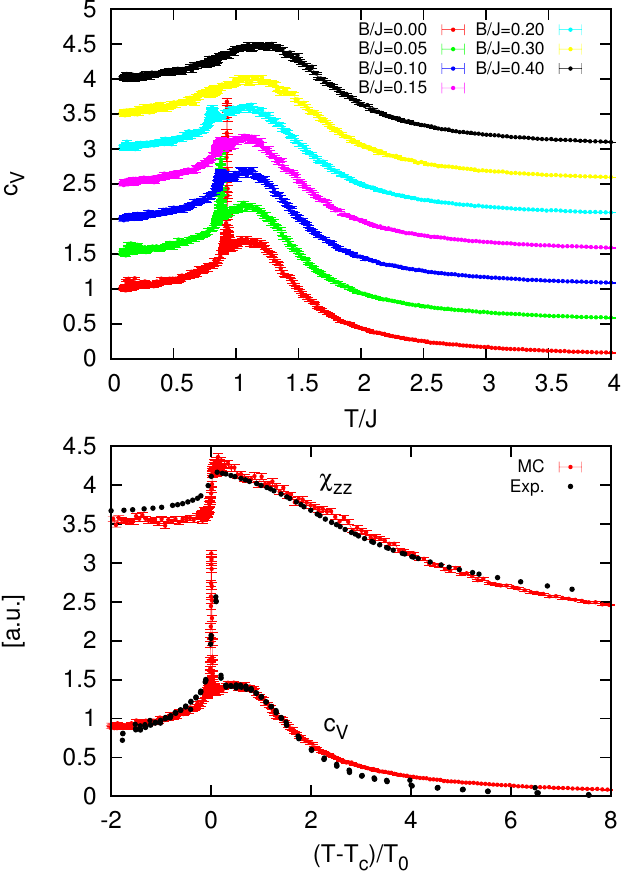}
	\caption{Specific heat and longitudinal susceptibility from Eq.~\eqref{eq:cv} for different magnetic fields and with anisotropy compensation according to Sec~\ref{sec:anisotropy}. Errorbars have been calculated using the jackknife method. An offset of 0.5 has been introduced to seperate different curves for the specific heat and $T_0$ was used as a fit parameter. The experimental data was taken from \cite{papermarkus}}
	\label{fig:td}
\end{figure}

\section{Thermodynamics across the temperature driven phase transition}
\label{sec:TD}
Besides the phase diagram we have studied the temperature driven phase transition into either the helical, conical, or skyrmion lattice phase (depending on the magnetic field) using PTMC.

The thermal transition at higher fields looks like a standard second order phase transition. At lower fields upon decreasing temperature there is an incipient behavior reminiscent of a second order phase transition. This behavior is controlled by the standard Wilson-Fisher fixed point. Upon decreasing temperature further below the Wilson-Fisher scale there is a second regime in which the system realizes that the critical modes do not go soft at a point in momentum space but instead on a whole sphere. Consequently, there is an abundance of soft modes which eventually drives the transition first order. The scenario outlined has been put forth by Brazovskii \cite{Brazovskii1975} and was recently studied in great detail in the context of MnSi~\cite{papermarkus}. The thermodynamic properties of the transition obtained from MC show striking similarity with the experimental findings~\cite{PhysRevB.76.052405,Pfleiderer200123,0953-8984-20-23-235222,Bauer2013,Lamago20051171} and the Brazovskii-scenario. We have studied two thermodynamic quantities, the specific heat $c_V$ and longitudinal susceptibility $\chi_{zz}$ ($\mathbf{B}~=~B\hat{\mathbf{z}}$). In MC both these quantities can be calculated from simple averages according to

\begin{equation}
	c_V(T) =  \frac{\langle E^2 \rangle - \langle E \rangle^2 }{N T^2}, \quad \chi_{zz}(T) =  \frac{\langle M_z^2 \rangle - \langle M_z \rangle^2 }{N T}.
	\label{eq:cv}
\end{equation}

Fig.~\ref{fig:td} shows the specific heat calculated from our simulations as well as a comparison of $\chi_{zz}$ to experimental data at zero field taken from~[\onlinecite{papermarkus}]. The specific heat for low fields clearly shows a first order peak on the low temperature side of the seeming second order transition. For higher fields there is a tendency of a vanishing first order peak which is an indication that the transition turns second order. As mentioned before, PTMC is not able to resolve the transition from the skyrmion to the conical phase and thus we do not observe any sign of this phase transition in our simulations. The longitudinal susceptibility compares well to the experimental data, in fact we find the characteristic drop of the susceptibility at $T_C$ as well as the characteristic inflection point at slightly higher temperatures\cite{Bauer2013}.
In all cases the MC data compares favorably to experiments. 

\section{Conclusion}
\label{sec:summary}

In this paper we have analyzed a full 3 dimensional classical spin model that describes chiral magnets like e.\,g. MnSi on a microscopic level. This effective model neglects the fact that these systems in general are no insulators but metallic in character. We used classical MC to determine the phase diagram and studied thermodynamic quantities across the thermal transition, such as the specific heat and the longitudinal susceptibility. From a simulation point of view, we identified the crucial role of lattice discretization anisotropies. After appropriate compensation we were able to reproduce the experimental phase diagram of MnSi qualitatively and have conclusively demonstrated that thermal fluctuations alone are sufficient to stabilize the skyrmion and that this assertion holds beyond Gaussian level. The calculated specific heat and longitudinal susceptibility shows remarkable agreement with the experimental data as well as recent analytical approaches.

We acknowledge discussions with M. Garst, C. Pfleiderer, A. Rosch, S. Trebst, and M. Vojta and thank A. Bauer for providing the experimental data. This work was supported by the Deutsche Forschungsgemeinschaft within the Emmy-Noether program through FR 2627/3-1 (SB,LF) and the Bonn-Cologne Graduate School for Physics and Astronomy (SB).


\begin{thebibliography}{10}%
\makeatletter
\providecommand \@ifxundefined [1]{%
 \ifx #1\undefined \expandafter \@firstoftwo
 \else \expandafter \@secondoftwo
\fi
}%
\providecommand \@ifnum [1]{%
 \ifnum #1\expandafter \@firstoftwo
 \else \expandafter \@secondoftwo
\fi
}%
\providecommand \enquote [1]{``#1''}%
\providecommand \bibnamefont  [1]{#1}%
\providecommand \bibfnamefont [1]{#1}%
\providecommand \citenamefont [1]{#1}%
\providecommand\href[0]{\@sanitize\@href}%
\providecommand\@href[1]{\endgroup\@@startlink{#1}\endgroup\@@href}%
\providecommand\@@href[1]{#1\@@endlink}%
\providecommand \@sanitize [0]{\begingroup\catcode`\&12\catcode`\#12\relax}%
\@ifxundefined \pdfoutput {\@firstoftwo}{%
 \@ifnum{\z@=\pdfoutput}{\@firstoftwo}{\@secondoftwo}%
}{%
 \providecommand\@@startlink[1]{\leavevmode}%
 \providecommand\@@endlink[0]{}%
}{%
 \providecommand\@@startlink[1]{%
  \leavevmode
  \pdfstartlink
   attr{/Border[0 0 1 ]/H/I/C[0 1 1]}%
   user{/Subtype/Link/A<</Type/Action/S/URI/URI(#1)>>}%
  \relax
 }%
 \providecommand\@@endlink[0]{\pdfendlink}%
}%
\providecommand \url  [0]{\begingroup\@sanitize \@url }%
\providecommand \@url [1]{\endgroup\@href {#1}{\urlprefix}}%
\providecommand \urlprefix [0]{URL }%
\providecommand \Eprint[0]{\href }%
\@ifxundefined \urlstyle {%
  \providecommand \doi [1]{doi:\discretionary{}{}{}#1}%
}{%
  \providecommand \doi [0]{doi:\discretionary{}{}{}\begingroup
  \urlstyle{rm}\Url }%
}%
\providecommand \doibase [0]{http://dx.doi.org/}%
\providecommand \Doi[1]{\href{\doibase#1}}%
\providecommand \bibAnnote [3]{%
  \BibitemShut{#1}%
  \begin{quotation}\noindent
    \textsc{Key:}\ #2\\\textsc{Annotation:}\ #3%
  \end{quotation}%
}%
\providecommand \bibAnnoteFile [2]{%
  \IfFileExists{#2}{\bibAnnote {#1} {#2} {\input{#2}}}{}%
}%
\providecommand \typeout [0]{\immediate \write \m@ne }%
\providecommand \selectlanguage [0]{\@gobble}%
\providecommand \bibinfo [0]{\@secondoftwo}%
\providecommand \bibfield [0]{\@secondoftwo}%
\providecommand \translation [1]{[#1]}%
\providecommand \BibitemOpen[0]{}%
\providecommand \bibitemStop [0]{}%
\providecommand \bibitemNoStop [0]{.\EOS\space}%
\providecommand \EOS [0]{\spacefactor3000\relax}%
\providecommand \BibitemShut [1]{\csname bibitem#1\endcsname}%
\bibitem{Muehlbauer}%
  \BibitemOpen
  \bibfield{author}{%
  \bibinfo {author} {\bibfnamefont{S.}~\bibnamefont{M\"uhlbauer}}
  \emph{et~al.},\ }%
  \bibfield{journal}{%
  \bibinfo {journal} {Science}\ }%
  \textbf{\bibinfo {volume} {323}},\ \bibinfo {pages} {915} (\bibinfo {year}
  {2009})%
  \bibAnnoteFile{NoStop}{Muehlbauer}%
\bibitem{Brazovskii1975}%
  \BibitemOpen
  \bibfield{author}{%
  \bibinfo {author} {\bibfnamefont{S.~A.}\ \bibnamefont{Brazovskii}},\ }%
  \bibfield{journal}{%
  \bibinfo {journal} {Sov. Phys. JETP}\ }%
  \textbf{\bibinfo {volume} {41}},\ \bibinfo {pages} {85} (\bibinfo {year}
  {1975})%
  \bibAnnoteFile{NoStop}{Brazovskii1975}%
\bibitem{papermarkus}%
  \BibitemOpen
  \bibfield{author}{%
  \bibinfo {author} {\bibfnamefont{M.}~\bibnamefont{Janoschek}} \emph{et~al.},\
  }%
  \bibfield{journal}{%
  \bibinfo {journal} {Phys. Rev. B}\ }%
  \textbf{\bibinfo {volume} {87}},\ \bibinfo {pages} {134407} (\bibinfo {year}
  {2013})%
  \bibAnnoteFile{NoStop}{papermarkus}%
\bibitem{Muenzer}%
  \BibitemOpen
  \bibfield{author}{%
  \bibinfo {author} {\bibfnamefont{W.}~\bibnamefont{M\"unzer}} \emph{et~al.},\
  }%
  \bibfield{journal}{%
  \bibinfo {journal} {Physical Review B}\ }%
  \textbf{\bibinfo {volume} {81}},\ \bibinfo {pages} {041203} (\bibinfo {year}
  {2010})%
  \bibAnnoteFile{NoStop}{Muenzer}%
\bibitem{Nature2DSkyrmions}%
  \BibitemOpen
  \bibfield{author}{%
  \bibinfo {author} {\bibfnamefont{X.}~\bibnamefont{Yu}} \emph{et~al.},\ }%
  \bibfield{journal}{%
  \bibinfo {journal} {Nature}\ }%
  \textbf{\bibinfo {volume} {465}},\ \bibinfo {pages} {901} (\bibinfo {year}
  {2010})%
  \bibAnnoteFile{NoStop}{Nature2DSkyrmions}%
\bibitem{NatureMaterials}%
  \BibitemOpen
  \bibfield{author}{%
  \bibinfo {author} {\bibfnamefont{X.}~\bibnamefont{Yu}} \emph{et~al.},\ }%
  \bibfield{journal}{%
  \bibinfo {journal} {Nature Material}\ }%
  \textbf{\bibinfo {volume} {10}},\ \bibinfo {pages} {106} (\bibinfo {year}
  {2011})%
  \bibAnnoteFile{NoStop}{NatureMaterials}%
\bibitem{Schulz12}%
  \BibitemOpen
  \bibfield{author}{%
  \bibinfo {author} {\bibfnamefont{T.}~\bibnamefont{Schulz}} \emph{et~al.},\ }%
  \bibfield{journal}{%
  \bibinfo {journal} {Nat. Phys.}\ }%
  \textbf{\bibinfo {volume} {8}},\ \bibinfo {pages} {301} (\bibinfo {year}
  {2012})%
  \bibAnnoteFile{NoStop}{Schulz12}%
\bibitem{Bogdanov1}%
  \BibitemOpen
  \bibfield{author}{%
  \bibinfo {author} {\bibfnamefont{A.}~\bibnamefont{Bogdanov}} \emph{et~al.},\
  }%
  \bibfield{journal}{%
  \bibinfo {journal} {J. Magn. Magn. Mater.}\ }%
  \textbf{\bibinfo {volume} {138}},\ \bibinfo {pages} {255} (\bibinfo {year}
  {1994})%
  \bibAnnoteFile{NoStop}{Bogdanov1}%
\bibitem{Bogdanov2}%
  \BibitemOpen
  \bibfield{author}{%
  \bibinfo {author} {\bibfnamefont{A.}~\bibnamefont{Bogdanov}} \emph{et~al.},\
  }%
  \bibfield{journal}{%
  \bibinfo {journal} {Sov. Phys. JETP}\ }%
  \textbf{\bibinfo {volume} {68}},\ \bibinfo {pages} {101} (\bibinfo {year}
  {1989})%
  \bibAnnoteFile{NoStop}{Bogdanov2}%
\bibitem{Tewari}%
  \BibitemOpen
  \bibfield{author}{%
  \bibinfo {author} {\bibfnamefont{S.}~\bibnamefont{Tewari}} \emph{et~al.},\ }%
  \bibfield{journal}{%
  \bibinfo {journal} {Phys. Rev. Lett.}\ }%
  \textbf{\bibinfo {volume} {956}},\ \bibinfo {pages} {047207} (\bibinfo {year}
  {2006})%
  \bibAnnoteFile{NoStop}{Tewari}%
\bibitem{Fischer}%
  \BibitemOpen
  \bibfield{author}{%
  \bibinfo {author} {\bibfnamefont{I.~A.}\ \bibnamefont{Fischer}}
  \emph{et~al.},\ }%
  \bibfield{journal}{%
  \bibinfo {journal} {Phys. Rev. B}\ }%
  \textbf{\bibinfo {volume} {77}},\ \bibinfo {pages} {024415} (\bibinfo {year}
  {2008})%
  \bibAnnoteFile{NoStop}{Fischer}%
\bibitem{Roessler}%
  \BibitemOpen
  \bibfield{author}{%
  \bibinfo {author} {\bibfnamefont{U.~K.}\ \bibnamefont{Roessler}}
  \emph{et~al.},\ }%
  \bibfield{journal}{%
  \bibinfo {journal} {Nature}\ }%
  \textbf{\bibinfo {volume} {442}},\ \bibinfo {pages} {797} (\bibinfo {year}
  {2006})%
  \bibAnnoteFile{NoStop}{Roessler}%
\bibitem{PhysRevB.80.054416}%
  \BibitemOpen
  \bibfield{author}{%
  \bibinfo {author} {\bibfnamefont{S.}~\bibnamefont{Yi}} \emph{et~al.},\ }%
  \bibfield{journal}{%
  \bibinfo {journal} {Phys. Rev. B}\ }%
  \textbf{\bibinfo {volume} {80}},\ \bibinfo {pages} {054416} (\bibinfo {year}
  {2009})%
  \bibAnnoteFile{NoStop}{PhysRevB.80.054416}%
\bibitem{0022-3719-13-31-002}%
  \BibitemOpen
  \bibfield{author}{%
  \bibinfo {author} {\bibfnamefont{P.}~\bibnamefont{Bak}} \emph{et~al.},\ }%
  \bibfield{journal}{%
  \bibinfo {journal} {J. Phys. C.}\ }%
  \textbf{\bibinfo {volume} {13}},\ \bibinfo {pages} {L881} (\bibinfo {year}
  {1980})%
  \bibAnnoteFile{NoStop}{0022-3719-13-31-002}%
\bibitem{JCC:JCC21450}%
  \BibitemOpen
  \bibfield{author}{%
  \bibinfo {author} {\bibfnamefont{C.~D.}\ \bibnamefont{Christ}}
  \emph{et~al.},\ }%
  \bibfield{journal}{%
  \bibinfo {journal} {J. Comput. Chem.}\ }%
  \textbf{\bibinfo {volume} {31}},\ \bibinfo {pages} {1569} (\bibinfo {year}
  {2010})%
  \bibAnnoteFile{NoStop}{JCC:JCC21450}%
\bibitem{PhysRevB.76.052405}%
  \BibitemOpen
  \bibfield{author}{%
  \bibinfo {author} {\bibfnamefont{S.~M.}\ \bibnamefont{Stishov}}
  \emph{et~al.},\ }%
  \bibfield{journal}{%
  \bibinfo {journal} {Phys. Rev. B}\ }%
  \textbf{\bibinfo {volume} {76}},\ \bibinfo {pages} {052405} (\bibinfo {year}
  {2007})%
  \bibAnnoteFile{NoStop}{PhysRevB.76.052405}%
\bibitem{Pfleiderer200123}%
  \BibitemOpen
  \bibfield{author}{%
  \bibinfo {author} {\bibfnamefont{C.}~\bibnamefont{Pfleiderer}}
  \emph{et~al.},\ }%
  \bibfield{journal}{%
  \bibinfo {journal} {J. Magn. Magn. Mater.}\ }%
  \textbf{\bibinfo {volume} {226-230, Pt. 1}},\ \bibinfo {pages} {23} (\bibinfo
  {year} {2001})%
  \bibAnnoteFile{NoStop}{Pfleiderer200123}%
\bibitem{0953-8984-20-23-235222}%
  \BibitemOpen
  \bibfield{author}{%
  \bibinfo {author} {\bibfnamefont{S.~M.}\ \bibnamefont{Stishov}}
  \emph{et~al.},\ }%
  \bibfield{journal}{%
  \bibinfo {journal} {J. Phys. C}\ }%
  \textbf{\bibinfo {volume} {20}},\ \bibinfo {pages} {235222} (\bibinfo {year}
  {2008})%
  \bibAnnoteFile{NoStop}{0953-8984-20-23-235222}%
\bibitem{Bauer2013}%
  \BibitemOpen
  \bibfield{author}{%
  \bibinfo {author} {\bibfnamefont{A.}~\bibnamefont{Bauer}} \emph{et~al.}}%
   (\bibinfo {year} {2013}),\
  \Eprint{http://arxiv.org/abs/1304.2407}{arXiv:1304.2407}%
  \bibAnnoteFile{NoStop}{Bauer2013}%
\bibitem{Lamago20051171}%
  \BibitemOpen
  \bibfield{author}{%
  \bibinfo {author} {\bibfnamefont{D.}~\bibnamefont{Lamago}} \emph{et~al.},\ }%
  \bibfield{journal}{%
  \bibinfo {journal} {Physica B}\ }%
  \textbf{\bibinfo {volume} {359-361}},\ \bibinfo {pages} {1171} (\bibinfo
  {year} {2005})%
  \bibAnnoteFile{NoStop}{Lamago20051171}%
\end{thebibliography}
\end{document}